\documentclass[runningheads]{llncs}

\usepackage{graphicx}
\usepackage[utf8]{inputenc} 
\usepackage[T1]{fontenc}    
\usepackage{hyperref}       
\usepackage{url}            
\usepackage{booktabs}       
\usepackage{amsfonts}       
\usepackage{nicefrac}       
\usepackage{microtype}      
\usepackage{threeparttable}
\usepackage{multirow}
\usepackage{amsmath}
\usepackage{color}
\usepackage{subfigure}
\usepackage{makecell}
\usepackage{multicol}
\usepackage{listings}
\usepackage{wrapfig}
\usepackage[linesnumbered, ruled]{algorithm2e} 
\usepackage{algorithmic}

\newenvironment{packeditemize}{
\begin{list}{$\bullet$}{
\setlength{\labelwidth}{8pt}
\setlength{\itemsep}{0pt}
\setlength{\leftmargin}{\labelwidth}
\addtolength{\leftmargin}{\labelsep}
\setlength{\parindent}{0pt}
\setlength{\listparindent}{\parindent}
\setlength{\parsep}{0pt}
\setlength{\topsep}{3pt}}}{\end{list}}

\begin{document}

\title{A Data Augmentation-based Defense Method Against Adversarial Attacks in Neural Networks}


\author{Yi Zeng\inst{1} \and
Han Qiu\inst{2} \and
Gerard Memmi\inst{2} \and
Meikang Qiu\inst{3}}

\authorrunning{Y. Zeng et al.}

\institute{University of California San Diego, CA, USA, 92122.\\
\email{y4zeng@eng.ucsd.edu}\\
\and
Telecom Paris, Institut Polytechnique de Paris, Palaiseau, France, 91120.\\
\email{\{han.qiu, gerard.memmi\}@telecom-paris.fr}\\
\and
Columbia University, New York, USA, 10027.\\
\email{qiumeikang@yahoo.com}
}

\maketitle 

\begin{abstract}

Deep Neural Networks~(DNNs) in Computer Vision~(CV) are well-known to be vulnerable to Adversarial Examples~(AEs), namely imperceptible perturbations added maliciously to cause wrong classification results. 
Such variability has been a potential risk for systems in real-life equipped DNNs as core components. Numerous efforts have been put into research on how to protect DNN models from being tackled by AEs. However, no previous work can efficiently reduce the effects caused by novel adversarial attacks and be compatible with real-life constraints at the same time. 
In this paper, we focus on developing a lightweight defense method that can efficiently invalidate full whitebox adversarial attacks with the compatibility of real-life constraints. 
From basic affine transformations, we integrate three transformations with randomized coefficients that fine-tuned respecting the amount of change to the defended sample. 
Comparing to 4 state-of-art defense methods published in top-tier AI conferences in the past two years, our method demonstrates outstanding robustness and efficiency. 
It is worth highlighting that, our model can withstand advanced adaptive attack, namely BPDA with 50 rounds, and still helps the target model maintain an accuracy around 80 \%, meanwhile constraining the attack success rate to almost zero. 

\end{abstract}

\footnotetext[1]{Han Qiu is the corresponding author.}

\section{Introduction}

With the rapid development of the Deep Neural Networks~(DNNs) in Computer Vision~(CV), there are more and more real-world applications that rely on the DNN models to classify images or to make decisions~\cite{lecun2015deep}. 
However, in recent years, the DNN models are well known to be vulnerable to {Adversarial Examples}~(AE) which threats the robustness of the DNN usage~\cite{szegedy2013intriguing}. 
Basically, the AEs can be generated by adding carefully designed perturbations that are imperceptible to human eyes but can mislead DNN classifiers with very high accuracy~\cite{goodfellow2014explaining}.

Today, several rounds of AE attack and corresponding defense techniques have been developed as shown in~\cite{qiu2020mitigating}. 
The initial research on adversarial attacks on DNN models such as Fast Gradient Sign Method (FGSM)~\cite{goodfellow2014explaining} aims at generating AEs by directly calculating the model gradients with respect to the input images. 
Such methods are then defeated by the defense methods based on various kinds of methods such as model distillation~\cite{papernot2016distillation}. 
Then, the improved AE attacks are proposed to combine the gradient-based approach with the optimization algorithm such as the CW~\cite{carlini2017towards} aims to find the input features that made the most significant changes to the final output to mislead the DNN models. 
Such an optimized gradient-based approach can defeat many previous defense methods including the model distillation. 
Later, advanced defense methods are proposed to mitigate such attacks by obfuscating the gradients of the inference process. 
Specifically, some data augmentation techniques are deployed in such defense methods such as image compression, image denoising, image transformation~\cite{qiu2020review}, etc. 
Then, such state-of-the-art methods are then defeated by more advanced attack methods such as Backward Pass Differentiable Approximation (BPDA)~\cite{athalye2018obfuscated} that can effectively approximate the obfuscated gradients to defeat these defenses. 

In this paper, we propose a novel defense method that combines several data augmentation techniques together to mitigate the adversarial attacks against on the DNN models. 
We propose our method, Stochastic Affine Transformation~(SAT), by deploying the image translation, image rotation, and image scaling method together.  
Our method can be used as a preprocessing step on the input images which makes our solution agnostic on many DNN models. 
Firstly, our method has little influence on the DNN inference which can effectively maintain the classification accuracy of benign images. 
Then, intensive experimentation and comparison have been performed to show the improvement of our method compared with several previous state-of-the-art defense solutions. 
Moreover, our method is a lightweight preprocess-only step that can be used on resource-constrained use cases such as the {Internet of Things}~(IoT)~\cite{qiu2020towards}. 
 
This paper includes two main contributions. (1) We design a data augmentation-based defense solution to mitigate the initial and optimized gradient-based adversarial attacks on DNN models. 
Our method combining several steps of data augmentation techniques can be used as a preprocessing step on input images that can effectively maintain the agnostic DNN model's accuracy. (2) Our method can also defeat the advanced adversarial attack method such as BPDA which outperforms many previous state-of-the-art defense solutions. 

This paper is organized as follows. 
Section \ref{sec:background} discusses the background information of this research including the brief definition of the adversarial examples and the previous data augmentation-based defense solutions.
Section \ref{sec:method} presents our threat model and defense requirements. 
Section \ref{sec:algorithm} proposes our methodology including the algorithm and the design details.
Section \ref{sec:eva} illustrates the experimentation details and evaluation results comparing with the previous state-of-the-art solutions. 
We then conclude in Section \ref{sec:con}. 

\newpage
\section{Research Backgrounds}
\label{sec:background}

In this section, we briefly introduce the background of the AEs in DNNs, the related work on AEs, and the state-of-the-art preprocessing-based defense methods based on data augmentation techniques.

\subsection{Adversarial Examples in Deep Neural Networks}

AEs can be explained as imperceptible modified samples that force one or multiple DNN models outputs with wrong results. This was first highlighted by~\cite{szegedy2013intriguing}. 
By denoting $I$ an input image, an adversarial example generated from it can be denoted as $\widetilde{I}=I+\delta$, where $\delta$ is the adversarial perturbation.
The target model, which conducts inference for classification tasks can be denoted as $f$, thus the problem of performing adversarial attacks on the target DNN model can be formulated as Eq~\ref{eq: ae1}. 

\begin{equation}
\label{eq: ae1}
    min  \lVert\delta\rVert ,\: s.t. \: f(\widetilde{I})\neq f(I)
\end{equation}

This equation can be interpreted as an optimization task that searches for a $\widetilde{I}$ based on $I$ that can be misclassified by the target model, while keeping $\widetilde{I}$ visually as similar to $I$ as possible. 
The aforementioned AE generation case is untargeted which aims to mislead the DNN classifier without a pre-set wrong label. 
As a targeted AE generation procedure aims to attack the DNN classifier to misclassify an input $I$ with original label $l$ as the pre-set wrong label $l'$. In the concern of real-life adoption of DNN models, both cases can result in serious outcomes if the models are not protected.

Since the time this vulnerability of DNNs has been discovered, various kinds of attacks have been proposed in the past few years to help the society to better understand the nature of AEs. 
To sum up, past work on adversarial attacks can be classified into two main approaches including {initial gradient-based} and {optimized gradient-based}. 
Fast Gradient Sign Method (FGSM)~\cite{goodfellow2014explaining} is one of the most famous initial gradient-based adversarial attacks which calculates the model gradients based on the sign of the gradient of the classification loss concerning the input image. 
FGSM performs a one-step gradient update along the direction of the sign of gradient at each pixel under $L_{\inf}$ constraints to generate AEs. 
Later on, variations of FGSM were introduced to better searching for the optimum AE based on a single input.
Such kind of methods includes I-FGSM~\cite{kurakin2016adversarial} and MI-FGSM~\cite{dong2017discovering}, aim at iteratively calculating the perturbations based on FGSM with a small step or with momentum. 

Then, optimized gradient-based AE attacks are proposed to calculate the gradients based on adopting optimization algorithms to find optimal adversarial perturbations directly between the input images and output predicted labels~\cite{carlini2017towards}. 
Such kind of attack is especially powerful in a whitebox or graybox scenario by adopting optimization algorithms to enhance the gradient calculation. 
Various optimized gradient-based AE attacks were proposed in recent years including Jacobian-based Saliency Map Attack (JSMA~\cite{papernot2016limitations}), DeepFool~\cite{moosavi2016deepfool}, LBFGS~\cite{szegedy2013intriguing},  Carlini \& Wagner (CW~\cite{carlini2017towards}), and Backward Pass Differentiable Approximation (BPDA~\cite{athalye2018obfuscated}).
We should highlight the BPDA attack here, as it invalidates dozens of existing state-of-art defense approaches in recent evaluations \cite{athalye2018obfuscated}.
The BPDA attack in a manner assumes that a defense function $g(\cdot)$ maintains the property $g(I)\approx I$ in order to preserve the functionality of the target model $f(\cdot)$. 
Then the adversary can use $g(I)$ on the forward pass and replace it with $I$ on the backward pass when calculating the gradients.

\subsection{Data Augmentation-based Preprocessing Defense Solutions} 

Various defensive strategies have been proposed to defeat adversarial attacks. 
One direction is to train a more robust model from either scratch or an existing model. 
Those approaches aim to rectify AEs' malicious features by including AEs into the training set~\cite{tramer2017ensemble}, processing all the training data~\cite{yang2019me}, or revising the DNN topology~\cite{papernot2016distillation}. 
However, training a DNN model is very time and resource-consuming, especially for real-life cases, where models are more complicated. 
Besides, in real-life, DNN models are packed as closed-source applications and cannot be modified, thus those methods are not applicable. Most of all, the adversary can still adaptively generate AEs for the new models~\cite{athalye2018obfuscated}. 

A more promising direction is to preprocess the input data to eliminate adversarial influence without touching the DNN model. 
These solutions are more suitable in the concern of real-life cases, as it is feasible, efficient, and lightweight. 
Thus, The preprocessing based defense is within the scope of this paper, as they do not require any laborious work with the DNN models, which made them competitive with most of the real-life defense scenarios.
Below we describe some previous works and their limitations:

\noindent{\textbf{Feature Distillation} (FD)~\cite{liu2019feature}} designed a compression method based on the JPEG compression but modified the quantization step. 
The basic idea is to measure the importance of input features for DNNs by leveraging the statistical frequency component analysis within the DCT of JPEG. 
It demonstrated a huge improvement in defending adversarial attacks compared with the standard JPEG compression method~\cite{qiu2020deep}. 

\noindent{\textbf{SHIELD}~\cite{das2018shield}} aims to randomize the quantization step by tuning the window size and quantization factors in the JPEG compression method. 
In SHIELD, the Stochastic Local Quantization (SLQ) method is used to divide an image into $8\times8$ blocks and applies a randomly selected JPEG compression quality (tuning quantization factors) to every block. 
The advantage is that the authors randomized the selective quantization steps which make the defense process different for different input images and make the adversarial attacks more difficult.

\noindent{\textbf{Bit-depth Reduction}~(BdR)~\cite{DBLP:conf/ndss/Xu0Q18}} performs a simple type of quantization that can remove small (adversarial) variations in pixel values from an image. In the evaluation of that work, it demonstrates a more effective result comparing to adversarial training. However, recently developed attacks are not within the scope of that work, namely the BPDA attack.

\noindent{\textbf{Pixel Deflection }~(PD)~\cite{prakash2018deflecting}} aims to add similar natural noises that are not sensitive to the DNN model. 
The idea of deflection is to randomly sample a pixel from an image and replace it with another randomly selected pixel from within a small square neighborhood. 
This could generate an artificial noise that affects little on the DNN model but can disturb the adversarial perturbations. 
Then, a BayesShrink denoising process is followed to recover the image content before this image is feed into the DNN model. 
The results of such a method are convincing since it introduces randomness into the preprocessing step and does not require any modification on the DNN model. 
However, the robustness of this method is significantly reduced if the attackers have knowledge of the preprocess step~\cite{athalye2018obfuscated}.

\section{Threat Model and Defense Requirements}
\label{sec:method}

\subsection{Threat Model}

Untargeted attacks and targeted attacks are two major types of adversarial attacks. Untargeted attacks try to mislead the DNN models to an arbitrary label different from the correct one. On the other hand, targeted attacks only considering succeed when the DNN model predicts the input as one specific label desired by the adversary \cite{carlini2017towards}. In this paper, we only evaluate the targeted attacks. The untargeted attacks can be mitigated in the same way. 

We consider a full whitebox scenario, where the adversary has full knowledge of the DNN model and the defense method, including the network architecture, exact values of parameters, hyper-parameters, and the details of the defense method. However, we assume the random numbers generated in real-time are perfect with a large entropy such that the adversary cannot obtain or guess the correct values. Such a targeted full whitebox scenario represents the strongest adversaries, as a big number of existing state-of-the-art defenses are invalidated as shown in~\cite{tramer2020adaptive}. 

As for the adversary's capability, we assume the adversary is outside of the DNN classification system, and he is not able to compromise the inference computation or the DNN model parameters (e.g., via fault injection to cause bit-flips \cite{rakin2019bit} or backdoor attacks \cite{gao2019strip}). What the adversary can do is to manipulate the input data with imperceptible perturbations. In the context of computer vision tasks, he can directly modify the input image pixel values within a certain range. We use $l_{\infty}$ and $l_{2}$ distortion metrics to measure the scale of added perturbations: we only allow the generated AEs to have either a maximum $l_{\infty}$ distance of 8/255 or a maximum $l_{2}$ distance of 0.05 as proposed in~\cite{athalye2018obfuscated}.

\subsection{Defense Requirements}

Various life-concerned vital tasks are already implemented with DNN in real-life, e.g., video surveillance~\cite{tang2017vehicle}, face authentication~\cite{mao2018privacy}, autonomous driving~\cite{wu2017squeezedet}, network traffic identification~\cite{zeng2019deep}, etc. Most of those cases' inference is conducted either locally in distributed computing units or remotely with the help of cloud servers.
In the case where inference procedures are conducted locally, resources are highly constrained, thus only lightweight designs of defense can protect the system without draw extra burden over those units.
Moreover, for both cases, previous resolutions, e.g. adversarial training or extra modifications over the target model would be considerably costly for real-life cases (where each sample is of large size, thus any kind of retraining can be laborious). As mentioned in the previous section, most of the existing defense methods either too complicated to be compatible with real-life constraints (\cite{buckman2018thermometer,guo2018countering,shaham2018understanding}) or not capable of effectively reduce the impact brought by adversarial attacks (\cite{prakash2018deflecting}).

To cope with those constraints in real-life adoption of DNNs, we believe the following properties should be taken considered when designing novel adversarial defense methods:
\begin{packeditemize}
\item {Accuracy-preserving}: they should not affect much on the prediction accuracy of the DNN model on clean data samples that processed by those methods. 
\item {Security}: they should be capable to effectively reduce effects brought by adversarial perturbations.
\item {Lightweight}: the defense method should not be too heavy to impact the devices' or units' performance or operations, considering the limited onboard computing capabilities and resources.
\item {Generalization Ability}: the defense method should neither require modifications over the target DNN's structure nor require any kind of retraining.
\end{packeditemize}

\begin{figure}
  \centering
  \includegraphics[width=0.8\linewidth]{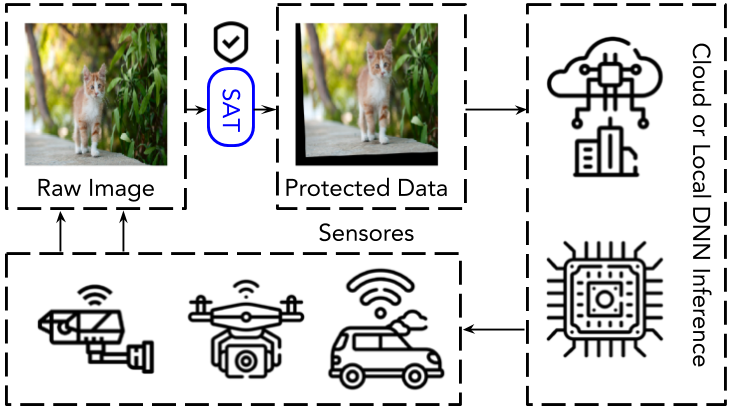}
  \caption{A system overview of adopting SAT in the real-life use case.}
  \label{fig:pro-tat}
\end{figure} 

Thus, for a better adaptation over nowadays real-life scenarios, we aim to design a preprocessing-only method to meet all those requirements. It has proved in our previous work \cite{qiu2020mitigating}, preprocessing-only adversarial defenses are competent enough to defense adversarial attacks, even for whitebox attacks.


\section{Proposed Methodology}
\label{sec:algorithm}

To better adapt to nowadays real-life scenarios' adoption of DNN, we present our efficient defense method against adversarial attacks that can conduct protections on the fly, termed {Stochastic Affine Transformation}~(SAT). Thanks to the lightweight design of SAT, this defense should be more compatible with both cloud DNN inference as well as local or edge DNN inference respecting real-life scenarios. \figurename~\ref{fig:pro-tat} illustrates an overview of adopting the SAT method in real life. The details of SAT will be present in Section \ref{PRO-TAT}. The analysis of the three hyperparameters respecting the defense efficiency is illustrated in Section \ref{PRO-TATHP}.

\subsection{SAT Algorithm}
\label{PRO-TAT}

Following the logic of adding randomness to the affine transformation without harming the classification accuracy \cite{qiu2020mitigating,guo2018countering,xie2018mitigating}, we propose a simple but effective way of image distortion as an adversarial example defense. 
The algorithm is designed based on combining several affine transformation methods.  
The details are illustrated in Algorithm 1.

Three basic affine transformations with randomized coefficients are bounded tother in this single procedure, namely, translation, rotation, and scaling. We add randomness to those three simple affine transformations so that the attacker cannot utilize a useful gradient to generate adversarial examples even acknowledges the details of this defense. Such is done by acquiring different coefficients that follow three uniform distribution for different samples. 
To be specific, there are three coefficients along with the raw image as the input of the SAT method. $T$ is the translation limit, $R$ is the rotation limit, and $S$ is the scaling limit.
The original input will first be randomly shifted away from its original coordinates according to $\delta_x$ and $\delta_y$ that both follow the uniform distribution in the range $(-T, T)$. Then, the data will be randomly rotated at a certain angle $\delta_r$ that follows the uniform distribution in the range $(-R, R)$. Finally, the distorted image will be acquired by scaling up or down $\delta_s$ times, where $\delta_s$ follows a uniform distribution in the range $(1-S,1+S)$. 

Since only simple affine transformations and random number generator are adopted in SAT, we believe SAT is compatible with both cloud DNN inference procedures as well as localized or edge devices DNN inference procedure. This lightweight design is also hardware friendly and can conduct protections on the fly.

\subsection{SAT Hyper-parameters}
\label{PRO-TATHP}

As aforementioned, there are three essential coefficients in the SAT method. In this part, we did a thorough evaluation and analysis of those three coefficients respecting the efficiency of defending adversarial attacks.

We believe a higher variance between the original data and the protected data while maintaining a high classification accuracy can help more to defend adversarial attacks, which is proved in our previous work \cite{qiu2020mitigating}. In this part, three different metrics are adopted to evaluate this variance, namely the $l_2$ norm, {Structural Similarity}~(SSIM) index, and the {Peak Signal-to-Noise Ratio}~(PSNR). The {classification accuracy}~(ACC) is the priority of most classification tasks thus is as well taken considerate in this part.

\SetKwInput{KwParam}{Parameters}
\begin{algorithm}[]
    \scriptsize
    \caption{{Stochastic Affine Transformation}}
    \label{algo:PRO-TAT}
    \SetNoFillComment
    \KwIn{original image $I\in \mathbb{R}^{h\times w}$}
    \KwOut{transformed image $I'\in \mathbb{R}^{h\times w}$}
    \KwParam{translation limit $T$; scaling limit $S$, rotation limit $R$.}
    \BlankLine
      $I^{'}=O^{h\times w}$\;
      \tcc{1.Translation}
      $\delta_x\sim\mathcal{U}(-T, T)$\; 
      $\delta_y\sim\mathcal{U}(-T, T)$\;
      $\Delta_x = \delta_x \times w$\; 
      $\Delta_y = \delta_y \times h$\;
      \If{$(x+\Delta_x \in (0, w)) \wedge (y+\Delta_y \in (0,h))$}{
          $I'(x,y)= I(x+\Delta_x,y+\Delta_y)$\; }
      \tcc{2.Rotation}
      $\delta_r\sim\mathcal{U}(-R, R)$\;
      $\Delta_r = \delta_r \times \pi/180$\;
      \For{$(x_i,y_j)$ in $\left \{ (x,y)|x\in(0,w),y\in(0,h)\right \}$}{
      $x_{i}^{'}=-(x_i-\left \lfloor w/2 \right \rfloor)\times sin(\Delta_r)+(y_j-\left \lfloor h/2 \right \rfloor)\times cos(\Delta_r)$\;
      $y_{j}^{'}=(x_i-\left \lfloor w/2 \right \rfloor)\times cos(\Delta_r)+(y_j-\left \lfloor h/2 \right \rfloor)\times sin(\Delta_r)$\;
      $x_{i}^{'}=\left \lfloor x_{i}^{'}+\left \lfloor w/2 \right \rfloor\right \rfloor$\;
      $y_{j}^{'}=\left \lfloor y_{j}^{'}+\left \lfloor h/2 \right \rfloor\right \rfloor$\;
      \If{$(x_{i}^{'} \in (0, w)) \wedge (y_{j}^{'} \in (0,h))$}{
          $I'(x_i,y_j)= I(x_{i}^{'},y_{j}^{'})$\; }
      }
      \tcc{3.Scaling}
      $\delta_s\sim\mathcal{U}(1-S, 1+S)$\;
      $h_{new} = \delta_s \times h$\;
      $w_{new} = \delta_s \times w$\;
      $I'= \texttt{reshape}(I',(h_{new},w_{new}))$\;
      \If{$\delta_s>1$}{
          $I'(x,y)= cropping(I',(h,w))$\; }
      \If{$\delta_s<1$}{
          $I'(x,y)= padding(I',(h,w))$\; }
      
      \Return $I'$\;
\end{algorithm}

$l_2$ is a widely adopted metric in deep learning domain to measure the amount of difference of two samples in the term of Euclidean distance, a higher $l_2$ indicates a greater difference. Eq. \ref{eq:l2nrom} shows how $l_2$ can be computed.

\begin{equation}
\label{eq:l2nrom}
    l_{2}(I',I) = \sqrt{(I_{R}'-I_{R})^2+(I_{G}'-I_{G})^2+(I_{B}'-I_{B})^2}/(h\times w\times 3)
\end{equation}
Where $I'$ and $I$ are the two 3-channel (RGB) samples to be compared. $h$ and $w$ are the height and width of those samples respectively.

SSIM is a metric in the computer vision domain normally being adopted to measure the similarity between two images, where smaller SSIM reflects greater difference. To be specific, SSIM is based on three comparison measurements between two samples, namely luminance ($l$), contrast ($c$), and structure ($s$). Each comparison function is elaborated in Eq \ref{eq:SSIM_l}, Eq \ref{eq:SSIM_c}, and Eq \ref{eq:SSIM_s} respectively.
\begin{equation}
\label{eq:SSIM_l}
    l(I',I)= \frac{2\mu_{I'}\mu_{I}+c_1}{\mu_{I'}^2+\mu_{I}^2+c_1}
\end{equation}
\begin{equation}
\label{eq:SSIM_c}
c(I',I)= \frac{2\sigma_{I'}\sigma_{I}+c_2}{\sigma_{I'}^2+\sigma_{I}^2+c_2}
\end{equation}
\begin{equation}
\label{eq:SSIM_s}
s(I',I)= \frac{\sigma_{I'I}+c_2/2}{\sigma_{I'}\sigma_{I}+c_2/2}
\end{equation}
Where $\mu(\cdot)$ computes the mean of a sample, $\sigma^2(\cdot)$ computes the variance of a sample. $c1$ is equal to $(0.01\times L)^2$, $c2$ is equal to $(0.03\times L)^2$. Here $L$ is the dynamic range of the pixel values. Finally, the SSIM can be acquired by computing the product of those three functions.

PSNR is most commonly used to measure the quality of reconstruction of lossy compression codecs, say compression, augmentation, or distortion, etc. A smaller PSNR indicates a greater difference between the two evaluating samples. PSNR can be defined via the mean squared error (MSE) between two comparing samples, which is explained in Eq \ref{eq:PSNR}.
\begin{equation}
\label{eq:PSNR}
PSNR(I', I)=20\cdot log_{10}(255)-10\cdot log_{10}(MSE(I',I))
\end{equation}
Where $MSE(\cdot)$ computes the MSE between two inputs.

\begin{wraptable}{R}{0.51\linewidth}
\centering
\caption{Comparisons of different methods over ACC and amount of changes.}
\newcommand{\tabincell}[2]{\begin{tabular}{@{}#1@{}}#2\end{tabular}}
  \small
\begin{tabular}{c c c c c}
\Xhline{1pt}
\textbf{Defense} & \textbf{$l_{2}$ norm} & \textbf{SSIM} & \textbf{PSNR}& \textbf{ACC}\\
\Xhline{1pt}
\tabincell{c} {SAT} & \textbf{0.322} & \textbf{0.194} & \textbf{10.219}& \textbf{0.98}\\
\hline 
\tabincell{c} {FD \cite{liu2019feature}} & 0.1343 & 0.4310 & 18.050 & 0.97\\
\hline 
\tabincell{c} {SHIELD \cite{das2018shield}} & 0.0405 & 0.8475 & 28.345 & 0.94\\
\hline 
\tabincell{c} {BdR \cite{DBLP:conf/ndss/Xu0Q18}} & 0.0709 & 0.7730 & 23.010 & 0.92\\
\hline 
\tabincell{c} {PD \cite{prakash2018deflecting}} & 0.0147 & 0.9877 & 37.100 & 0.97\\
\hline 
\Xhline{1pt}
\end{tabular}
\label{tab: compare}
\end{wraptable}

\begin{figure*}[!htbp]
\centering
\subfigure[ACC as a function over $T$,$S$,$R$]
{\includegraphics[width = 0.48\linewidth]{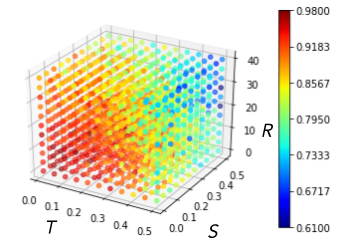}\label{PRO-TAT_ACC}}
\subfigure[$l_2$ as a function over $T$,$S$,$R$] 
{\includegraphics[width = 0.48\linewidth]{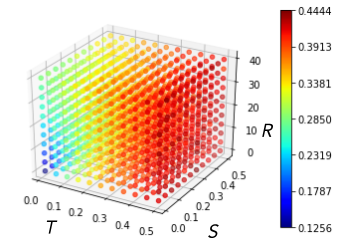}\label{PRO-TAT_L2}}
\subfigure[SSIM as a function over $T$,$S$,$R$] 
{\includegraphics[width = 0.48\linewidth]{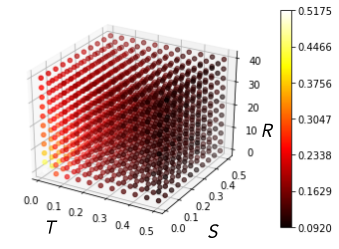}\label{PRO-TAT_SSIM}}
\subfigure[PSNR as a function over $T$,$S$,$R$] 
{\includegraphics[width = 0.48\linewidth]{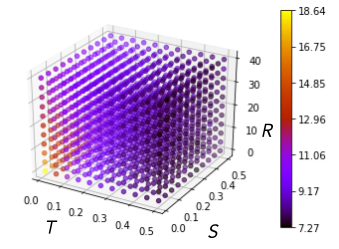}\label{PRO-TAT_PSNR}}
\caption{Metrics of reconstructed images under different values of $T$,$S$,$R$}
\label{PRO-TATmetrics}
\end{figure*}

We tried different values of $T$ and $S$ in the range $[0.01,0.5]$. Different values of $R$ is acquired in the range $[0,40]$. Thus, each coefficient will test 11 values in the respecting range. As for different metrics, we will acquire 1331 $(11\times11\times11)$ results from different combinations of those different values of coefficients.
\figurename~\ref{PRO-TATmetrics} demonstrates the change of those four metrics' value when different $T$, $S$,and $R$ are adopted.

In \figurename~\ref{PRO-TAT_ACC}, the changes of ACC with those three coefficients varies is presented. The right side of \figurename~\ref{PRO-TAT_ACC} is the color-bar that reflecting the ACC attained with respecting combinations of those three coefficients. We can learn that lower $T$, $S$, and $R$ can help the model maintain a high ACC. To ensure a high ACC, we set 95\% ACC as a standard, thus those combinations with ACC below this standard would not be taken further considerations.

From \figurename~\ref{PRO-TAT_L2} we can learn that the $l_2$ is not that sensitive with $S$ and $R$ comparing to $T$ in their respecting range. Combining the information provided from \figurename~\ref{PRO-TAT_L2}, \figurename~\ref{PRO-TAT_SSIM}, and \figurename~\ref{PRO-TAT_PSNR} we can as well acquire this similar analytical result for SSIM and PSNR. This phenomenon also reflects that the $l_2$, SSIM, and PSNR can all reflecting the scale of changes in a similar manner respecting affine transformations.

By overlapping \figurename~\ref{PRO-TAT_ACC} with the other three figures, we can acquire a set of optimum coefficients that ensures high ACC and great variance at the same time. The set of coefficients for the following experiment is set as follows: $T=0.16$, $S=0.16$, and finally $R=4$.

We compare the SAT method using those fine-tuned hyperparameters with other state-of-art adversarial defense methods, which presented in~\tablename~\ref{tab: compare}. As demonstrated, SAT can create a greater difference between raw samples and protected samples while maintaining a high ACC than other methods. We will evaluate whether this greater variance will help and how much will it help DNN models to defend adversarial attacks in the following section.

\section{Experimentation and Evaluation}
\label{sec:eva}

In this section, we conduct a comprehensive evaluation of the proposed technique. Various adversarial attacks are taken considered in this part: 4 kinds of standard adversarial attacks (FGSM, I-FGSM, LBFGS, and C\&W) are conducted, advanced interactive gradient approximation attack, namely BPDA, is also conducted to evaluate the robustness of SAT. We compare SAT with four state-of-art defense methods publish in top-tier artificial intelligence conferences from the past two years. This section would be divided into three parts to elaborate on the settings of the experiment, efficiency over standard adversarial attacks, and the efficiency over BPDA respectively.

\subsection{Experimental Settings}
Tensorflow~\cite{abadi2016tensorflow} is adopted as the deep learning framework to implement the attacks and defenses. The learning rate of the C\&W and BPDA attack is set to 0.1. 
All the experiments were conducted on a server equipped with 8 Intel I7-7700k CPUs and 4 NVIDIA GeForce GTX 1080 Ti GPU.

SAT is of general-purpose and can be applied to various models over various platforms as a preprocessing step for computer vision tasks as illustrated in \figurename~\ref{fig:pro-tat}. Without the loss of generality, we choose a pre-trained Inception V3 model \cite{szegedy2016rethinking} over the ImageNet dataset as the target model. This state-of-the-art model can reach 78.0\% top-1 and 93.9\% top-5 accuracy. We randomly select 100 images from the ImageNet Validation dataset for AE generation. These images can be predicted correctly by this Inception V3 model. 

We consider the targeted attacks where each target label different from the correct one is randomly generated \cite{athalye2018obfuscated}. 
For each different attack, we measure the classification accuracy of the generated AEs (ACC) and the attack success rate (ASR) of the targeted attack.
To be noticed that the untargeted attacks are not within our scope in this work. A higher ACC or lower ASR indicates the defense is more resilient against the attacks. 

For comparison, we re-implemented 4 existing solutions including FD~\cite{liu2019feature}, SHIELD~\cite{das2018shield}, Bit-depth Reduction~\cite{DBLP:conf/ndss/Xu0Q18}, and PD~\cite{prakash2018deflecting}. 

\subsection{Evaluation on Defending Adversarial Attacks}

We first evaluate the efficiency of our proposed method over standard adversarial attacks, namely FGSM, I-FGSM, C\&W, and LBFGS. For FGSM and I-FGSM, AEs are generated under $l_{\infty}$ constraint of 0.03. 
For LBFGS and C\&W, the attack process is iterated under $l_2$ constraint and stops when all targeted AEs are found. 
We measure the model accuracy (ACC) and attack success rate (ASR) with the protection of SAT and other defense methods.

\begin{table}
\centering
\caption{Comparisons of different defense against attacks respecting ACC.}
\newcommand{\tabincell}[2]{\begin{tabular}{@{}#1@{}}#2\end{tabular}}
  \small
\begin{tabular}{c|c|c|c|c|c}
\Xhline{1pt}
\textbf{Defense} & \textbf{Clean} & \textbf{FGSM($\epsilon$=.03)} & \textbf{IFGSM($\epsilon$=.03)} & \textbf{C\&W} & \textbf{LBFGS} \\
\Xhline{1pt}
\tabincell{c} {{Baseline}} & {1.00} & {0.42} & {0.02} & {0.06} & {0.00}\\
\hline 
\tabincell{c} {SAT} & \textbf{0.98} & \textbf{0.61} & \textbf{0.85} & \textbf{0.78}& \textbf{0.96}\\
\hline 
\tabincell{c} {FD \cite{liu2019feature}} & 0.97 & 0.47 & \textbf{0.87} & \textbf{0.84} & \textbf{0.97} \\
\hline 
\tabincell{c} {SHIELD \cite{das2018shield}} & 0.94 & 0.49 & 0.84 & \textbf{0.78} & 0.92\\
\hline 
\tabincell{c} {BdR \cite{DBLP:conf/ndss/Xu0Q18}} & 0.92 & 0.47 & 0.82 & 0.61 & 0.90\\
\hline 
\tabincell{c} {PD \cite{prakash2018deflecting}} & 0.97 & 0.42 & 0.30 & 0.11 & 0.86\\
\Xhline{1pt}
\end{tabular}
\label{tab: ACC}
\end{table}


The results are shown in~\tablename~\ref{tab: ACC} and~\tablename~\ref{tab: ASR}. 
For benign samples only, our proposed techniques have the smallest influence on the model accuracy comparing to past works. 
For defeating AEs generated by these standard attacks, the attack success rate can be kept around 0\% and the model accuracy can be drastically recovered, which an efficiency against different kinds of adversarial attacks is demonstrated. 
To be noticed, previous work can only attain an accuracy of around 50\% on samples attacked by the FGSM ($\epsilon = .03$). SAT can recover the accuracy to 0.61\%.

Comparing to the~\tablename~\ref{tab: compare} which compares different methods' capability of creating a variance between input and defended sample, the defense efficiency evaluated in this part shows a strong correlation with the amount of variance generated by the defense method. This has confirmed the previous conclusion in our previous work \cite{qiu2020mitigating}. 

In a nutshell, the effectiveness of SAT against standard adversarial attack is demonstrated, as we can attain a state-of-art defense efficiency on all the evaluated attacks comparing to other methods.

\begin{table}
\centering
\caption{Comparisons of different defense against attacks respecting ASR.}
\newcommand{\tabincell}[2]{\begin{tabular}{@{}#1@{}}#2\end{tabular}}
  \small
\begin{tabular}{c|c|c|c}
\Xhline{1pt}
\textbf{Defense}   & \textbf{IFGSM($\epsilon$=.03)} & \textbf{C\&W} & \textbf{LBFGS} \\
\Xhline{1pt}
\tabincell{c} {{Baseline}}   & {0.95} & {1.00} & {1.00}\\
\hline 
\tabincell{c} {SAT}   & \textbf{0.01} & \textbf{0.01} & \textbf{0.00}\\
\hline 
\tabincell{c} {FD \cite{liu2019feature}}   & \textbf{0.00} & \textbf{0.06} & \textbf{0.00}\\
\hline 
\tabincell{c} {SHIELD \cite{das2018shield}}   & \textbf{0.01} & 0.02 & \textbf{0.00}\\
\hline 
\tabincell{c} {BdR \cite{DBLP:conf/ndss/Xu0Q18}}   & 0.02 & 0.2 & \textbf{0.00}\\
\hline 
\tabincell{c} {PD \cite{prakash2018deflecting}}   & 0.6 & 0.11 & 0.02\\
\Xhline{1pt}
\end{tabular}
\label{tab: ASR}
\end{table}

\footnotetext[2]{The ASR evaluation of the FGSM is not shown here since the baseline ASR is 0.}

\subsection{Evaluation on Defending Advanced Adversarial Attacks}

We then evaluate the effectiveness of SAT against the BPDA attack. 
Since the BPDA attack is an interactive attack, we record the ACC and ASR for each round for different defense methods.

\begin{figure*}[!htbp]
\centering
\subfigure[ACC per round under BPDA attack.]
{\includegraphics[width = 0.49\linewidth]{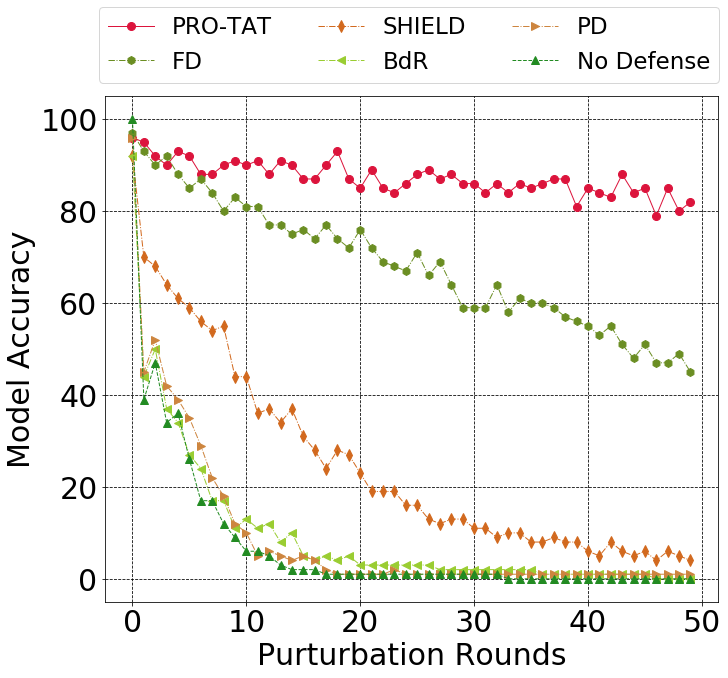}\label{BPDA_acc}}
\subfigure[ASR per round under BPDA attack.] 
{\includegraphics[width = 0.49\linewidth]{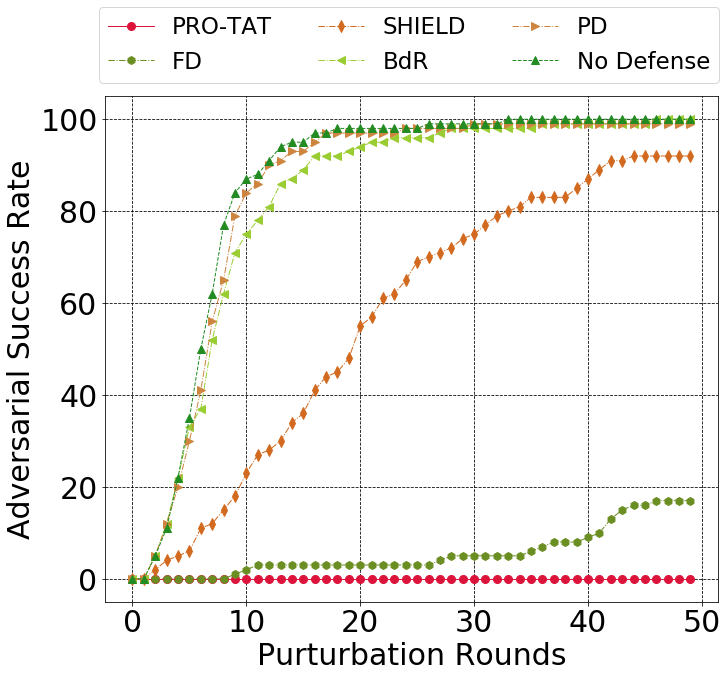}\label{BPDA_suc}}
\caption{ACC and ASR of various techniques under the BPDA attack. }
\label{Evaluation}
\end{figure*}

The model prediction accuracy and attack success rate in each round are shown in \figurename~\ref{BPDA_acc} and \ref{BPDA_suc}, respectively. We can observe that after 50 attack rounds, all other three prior solutions except FD can only keep the model accuracy lower than 5\%, and attack success rates reach higher than 90\%. 
Those defenses fail to mitigate the BPDA attack.
FD can keep the attack success rate lower than 20\% and the model accuracy is around 40\%. This is better but still not very effective in maintaining the DNN model's robustness.

In contrast, SAT is particularly effective against the BPDA attack. 
As our method can maintain an acceptable model accuracy (around 80\% for 50 perturbation rounds), and restrict the attack success rate to 0 for all the record rounds. 
This result is as well consistent with the $l_2$, SSIM, and PSNR metrics compared in \tablename~\ref{tab: compare}: the randomization effects in SAT cause greater variances between $I'$ and $I$, thus invalidating the BPDA attack basic assumption, which is $I' \approx I$. 

To sum up, the effectiveness of SAT against the BPDA attack is demonstrated, as a considerable improvement over the defense efficiency against the BPDA attack is shown comparing to previous work.

\section{Conclusion}
\label{sec:con}
In this paper, we proposed a lightweight defense method that can effectively invalidate adversarial attacks, termed SAT. By adding randomness to the coefficients, we integrated three basic affine transformations into SAT. Compared with four state-of-art defense methods published in the past two years, our method clearly demonstrated a more robust and effective defense result on standard adversarial attacks. Moreover, respecting the advanced BPDA attack, SAT showed an outstanding capability of maintaining the target model's ACC and detain the ASR to 0. This result is almost 50\% better than the best result achieved by previous work against full whitebox targeted attacks.

\bibliographystyle{splncs04}
\bibliography{ref}

\end{document}